# Thermodynamic Stabilization of Mixed-Halide Perovskites Against Phase Segregation

Eline M. Hutter[1#], Loreta A. Muscarella[1], Francesca Wittmann[1], Jan Versluis[1], Lucie McGovern[1], Huib J. Bakker[1], Young-Won Woo[2], Young-Kwang Jung[2], Aron Walsh[2,3], Bruno Ehrler[1]

1. AMOLF, Science Park 104, 1098 XG Amsterdam, the Netherlands
2. Department of Materials Science and Engineering, Yonsei University, Seoul 03722, Korea
3. Department of Materials, Imperial College London, London SW7 2AZ, United Kingdom
#. Current address: Department of Chemistry, Utrecht University, Princetonlaan 8, 3584 CB, Utrecht, the Netherlands

Lead contact:

Dr. Eline Hutter: e.m.hutter@uu.nl

Corresponding author:

Dr. Bruno Ehrler: b.ehrler@amolf.nl

**Abstract**

Mixing iodide and bromide in halide perovskite semiconductors is an effective strategy to tune their bandgap, therefore mixed-halide perovskites hold great promise for color-tunable LEDs and tandem solar cells. However, the bandgap of mixed-halide perovskites is unstable under (sun-)light, since the halides segregate into domains of different bandgaps. Using pressure-dependent ultrafast transient absorption spectroscopy, we show that high external pressure increases the range of stable halide mixing ratios. Chemical compression, by inserting a smaller cation, has the same effect, which means that any iodide-to-bromide ratio can be stabilized by tuning the crystal volume and compressibility. We interpret these findings as an increased thermodynamic stabilization through alteration of the Gibbs free energy via the largely overlooked $P\Delta V$ term.

**Introduction**

Metal halide perovskite semiconductors have recently received tremendous attention in materials science, as these have yielded highly efficient solar cells, light-emitting diodes (LEDs), and radiation detectors.[1–3] The unprecedented performance of perovskites is due to their outstanding optoelectronic properties, such as high optical absorption coefficients and relatively low trap densities, resulting in excellent charge transport and efficient radiative recombination.[4,5] Another key characteristic of metal-halide perovskites is that their bandgap is highly dependent on their chemical composition, meaning that any desired absorption onset or emission energy in the visible can be obtained by tuning the composition. For instance, mixing iodide and bromide in $MAPb(I_{1-x}Br_x)_3$ (with MA = methylammonium, $CH_3NH_3^+$) results in bandgaps intermediate to full iodide ($x = 0$, 1.6 eV) and full bromide ($x = 1$, 2.3 eV).[6,7] The combination of excellent optoelectronic properties and bandgap tunability makes perovskites the most promising candidate for color-tunable LEDs as well as tandem solar cells,[8] where multiple semiconductors with different bandgaps are used to achieve optimum power conversion efficiencies.[9]

A major drawback of mixed-halide perovskites is that their bandgap is unstable under illumination, since the halides segregate into iodide-rich and bromide-rich domains.[10,11] The lower-bandgap iodide-rich domains act as charge carrier recombination centers and this halide segregation is thus detrimental for device performance. It is thus essential that the halide segregation is fully suppressed for any application of mixed-halide perovskites where stable bandgaps are required.

Previous studies have shown that halide segregation can be retarded by reducing the defect density and/or the illumination intensity.[12] However, achieving a commercially relevant stability on the timescale of 20 years may not be feasible through retardation alone (kinetic stability), as this would require a five orders of magnitude reduction in segregation rate.[13] Unravelling design rules for *thermodynamically* stable mixed-halide perovskites will thus be vital for bandgap-tunable perovskites with long-term stability. At ambient conditions, segregation of $MAPb(I_{1-x}Br_x)_3$ terminates at $x = 0.2$,[10,12] and segregation is absent if $x < 0.2$, which was attributed to a thermodynamic minimum in the Gibbs free energy.[14–16] In this work, we show that both external pressure and chemical compression shift this terminal *x*-value (up to ~0.6) and consequently, the range of thermodynamically stable mixing ratios is significantly extended under compression.[17] We extracted these terminal *x*-values from transient absorption spectroscopy (TAS) measurements performed at hydrostatic pressures ranging from ambient to 0.3 GPa and for several initial mixing ratios *x*. In contrast to previously reported photoluminescence measurements at high pressure,[11] TAS allows to track the formation of *both* the iodide- and bromide-rich domains during segregation. Hence, we find that at high pressure, both the iodide- and bromide-rich phase are closer to the initial *x*, and the terminal *x*-value depends on both the external pressure and the initial composition. These findings can be understood from a change in the thermodynamics (Gibbs free energy) due to changes in the compressibility and the unit cell volume, providing an effective approach to obtain mixed-halide perovskites that are stable against photo-induced halide segregation. Consistently, we find that chemical compression of the perovskite, *via* replacing MA ions with the smaller Cs ions, effectively suppresses halide segregation at ambient pressure. Altogether, our results show that stable mixed-halide perovskites of any desired halide composition can be designed by modifying the mechanical properties of the crystal so that the desired halide ratio falls in a miscible regime of the phase diagram, enabling a rational route toward thermodynamically stable mixed-halide perovskites.

**Results and discussion**

The MAPb($I_{1-x}Br_x$)$_3$ perovskite is an example of a pseudo-binary mixture (or solid solution). The accessible solubility range is determined by the free energy of mixing $\Delta G(x)$: the free energy of the mixed phase with respect to the phase-separated iodide ($x = 0$) and bromide ($x = 1$) compounds:

$$\Delta G(x) = \Delta H(x) - T\Delta S(x) \qquad (1)$$

First-principles calculations of Equation 1 confirm a positive enthalpic term $\Delta H(x = 0.5) \sim 2$ kJ/mol due to chemical strain in the mixed MAPb($I_{1-x}Br_x$)$_3$ perovskite,[14] which originates from the ionic size mismatch of I$^-$ (2.22 Å) and Br$^-$ (1.96 Å). This enthalpic cost of straining the bonding environment is offset by the gain in configurational entropy ($T\Delta S$).[14] For a binary mixture, the entropy reaches a maximum at $x = 0.5$ with a value of also ~ 2 kJ/mol (0.7 $k_BT$ at $T$ = 300 K). Analysis of $\Delta G(x)$ shows two minima (at $x$ ~ 0.2 and at 0.75) under ambient conditions. The region in between these minima represents the miscibility gap (*i.e.* the range of compositions that are thermodynamically unstable). What has been overlooked thus far is that the enthalpy contains a $P\Delta V$ term where pressure ($P$) can be used as an additional lever to control the stability range, which we explore in the following.

Thin films of mixed-halide $MAPb(I_{1-x}Br_x)_3$ with $0 < x < 1$ were spin-coated from solution, see Experimental Methods for procedures and Supporting Information Figure 1 for UV-VIS and XRD, and their initial stoichiometries $x$ were determined from SEM-EDX. The initial bandgaps vary from 1.6 eV ($x$ = 0) to 2.3 eV ($x$ = 1). However, the bandgaps of $0.2 < x < 1$ are unstable under illumination with a continuous wave (cw) laser, showing photoluminescence (PL) emission energies of 1.68 – 1.75 eV, independent of $x$ (see Supporting Information Figure 2). This bandgap instability has been widely observed in mixed-halide perovskites,[10,18–20] and is attributed to segregation of the halides into iodide- and bromide-rich domains. After segregation, all light emission originates from the iodide-rich domains, which have a lower bandgap. Therefore, this halide segregation is detrimental for perovskite-based LEDs if their desired emission energy falls in between 1.7 and 2.3 eV. Segregation is also unfavorable for perovskite-based solar cells, as the voltage is lowered by $\Delta E$: the difference between the original bandgap and the energy of the iodide-rich region.[19] The shared emission energy of ~ 1.7 eV corresponds to a segregated composition $x_s$ ~ 0.2, and the $MAPb(I_{1-x}Br_x)_3$ perovskite is unstable if $x > x_s$ (= 0.2 at ambient conditions).[10,21] Since this threshold coincides with a cubic-to-tetragonal phase transition around room temperature, it was previously proposed that this phase transition impedes full halide segregation.[20,21] As we show below, however, this assignment is inconsistent with the behavior of mixed-halide perovskites under hydrostatic pressure.

We used pressure-dependent TAS measurements, as shown in Figure 1a, to follow the segregation in time for different initial mixing ratios and pressures. Importantly, while PL measurements only probe emissive phases, and thus often only probe the iodide-rich phase,[10,17] TAS measures the bleach from each excited state population and hence allows us to get a full picture on the energetic landscape, tracing both the formation of iodide-rich and bromide-rich phases.[22,23] We aligned the pump and probe inside a hydraulic pressure cell, filled with the perovskite thin film and an inert liquid (degassed perfluorohexane, see Experimental Methods). Increasing the liquid content in the cell increases the hydrostatic pressure up to 0.3 GPa. We note that this pressure range is low enough to avoid any phase transitions in the MAPb($I_{1-x}Br_x$)$_3$ perovskites, as confirmed by pressure-dependent UV-VIS measurements (see Supporting Information Figure 3).

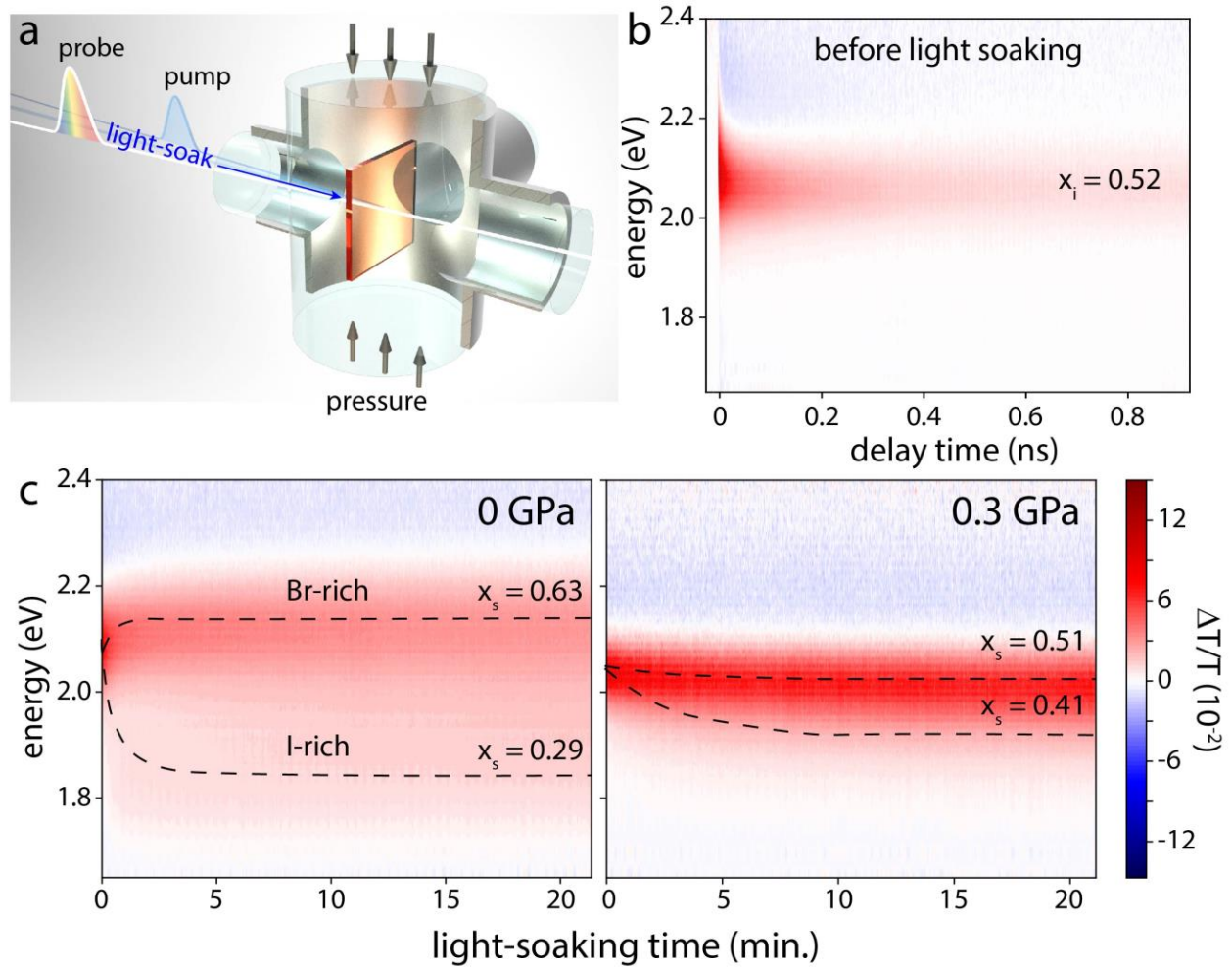

**Figure 1: Pressure-dependent transient absorption spectroscopy (TAS) of MAPb($I_{1-x}Br_x$)$_3$ thin film before and after light soaking.** a) Schematic representation of the TAS set-up, where the MAPb($I_{1-x}Br_x$)$_3$ thin film is placed inside a hydraulic pressure cell. The pressure can be increased by adding more liquid, and the pump, probe and cw-laser enter this cell via quartz windows. b) $\Delta T/T$ as a function of energy and delay time for $x = 0.52 \pm 0.04$ before light soaking. (c-d) $\Delta T/T$ as a function of energy during 20 minutes of light soaking with the cw-laser ($\lambda = 405$ nm, $I = 2.37 \times 10^3$ mW/cm$^2$) at 0 GPa (c) and 0.3 GPa (d). At ambient pressure

(c), light soaking leads to segregation into iodide-rich ($x_s = 0.29 \pm 0.02$) and bromide-rich ($x_s = 0.63 \pm 0.05$) domains, while light soaking at 0.3 GPa only leads to the formation of a small side peak corresponding to $x_s = 0.41 \pm 0.03$.

Figure 1b shows $\Delta T/T$ as a function of energy and delay time $t_d$ (time after excitation) for a thin MAPb(I$_{1-x}$Br$_x$)$_3$ film with $x = 0.52 \pm 0.04$. The positive (bleach) signal peaking at 2.02 eV corresponds to the ground state bleach of the perfectly mixed iodide-bromide perovskite. The recombination of these charges then leads to a reduction in $\Delta T/T$ at increased $t_d$.

To investigate the effect of light soaking on the energetic landscape of the perovskite, we illuminated the sample with a continuous-wave (cw, intensity equivalent to 24 suns) laser for 20 minutes. This relatively high light-soaking intensity was used so that we could perform the measurement within a reasonable timeframe, especially at the higher pressures. A fixed delay time of ~15 ps was used, which is prior to recombination and energy transfer events but after cooling of the charges into the iodide- and bromide-rich domains.[23,24] As shown in Figure 1c, two positive features appear within several minutes of light soaking, indicating the segregation into iodide- and bromide-rich domains, with bandgaps of 1.84 and 2.11 eV, respectively (see also Supporting Information Figure 4). These energies correspond to $x_s$-values of $0.29 \pm 0.02$ for the iodide-rich phase, and $x_s = 0.63 \pm 0.05$ for the bromide-rich phase (see Supporting Information Figure 5). In contrast, light soaking the same mixed-halide perovskite at high pressure (0.3 GPa) does not lead to the formation of distinct iodide- and bromide-phases. Instead, only a small side peak appears corresponding to $x_s = 0.41 \pm 0.03$, next to the initial $\Delta T/T$ peak ($x_s = 0.51 \pm 0.04$). This observation shows that halide segregation is substantially suppressed at high pressure.

Figure 2a-c show $\Delta T/T$ during light soaking at ambient pressure for mixed-halide perovskites with $x = 0.25$ (a), $x = 0.5$ (b) and $x = 0.7$ (c). The initial GSBs peaking at 2.1 eV (at 15 ps) for $x = 0.5$ (Figure 2b) and 2.2 eV for $x = 0.7$ (Figure 2c) shift to higher energies during light soaking, while reducing in magnitude. Simultaneously, an ingrowth of a relatively broad peak is observed around 1.9 eV, due to the formation of the iodide-rich phase. For $x = 0.25$, the GSB slightly red-shifts 30 meV during light soaking, without the formation of a (detectable) second phase, see Figure 2a. Similarly, no phase segregation is observed on light

soaking for $x = 0.1$, see Supporting Information Figure 6, consistent with previously reported PL measurements showing stable bandgaps for $x < 0.2$.[21]

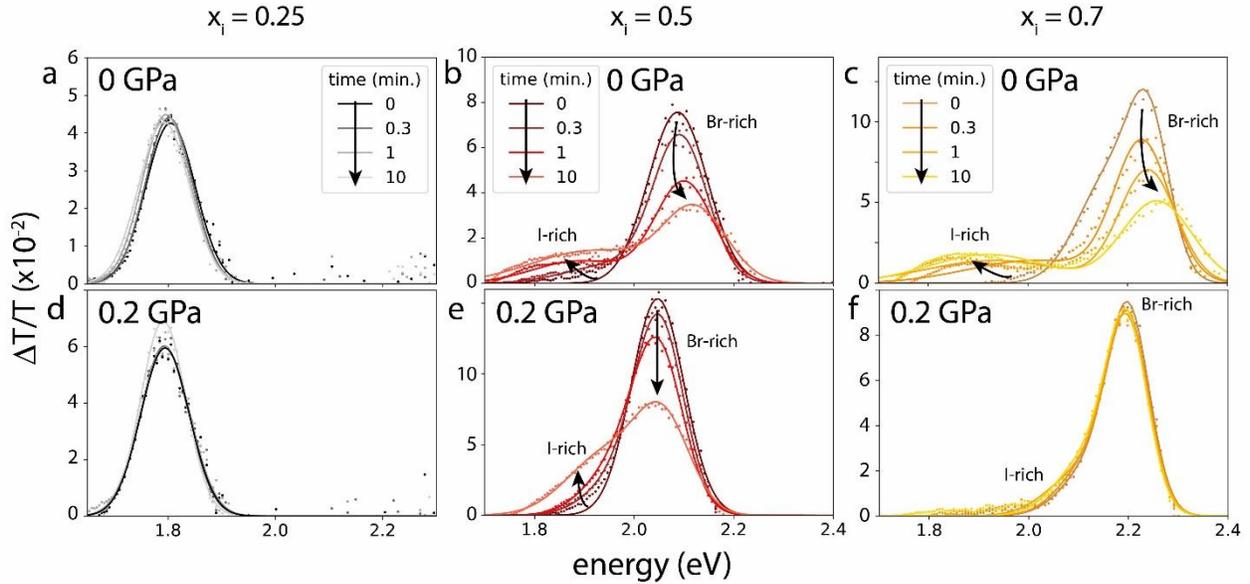

**Figure 2**. (a-f) Transient absorption spectra (recorded at 15 ps time delay) after 0, 0.3, 1 and 10 minutes of light soaking at ambient pressure using cw illumination (405 nm, 2.37 x $10^3$ mW/cm²) for $x = 0.25$ (a), $x = 0.5$, (b) and $x = 0.7$ (c) and at high pressure (d to f). Note that (b) and (e) correspond to the sample shown in Figure 1.

While the iodide-rich phase has the same final composition independent of initial composition (at ambient pressure), we find that the bromide-rich phase changes substantially with $x$, as indicated by the different positions of the high-energy peaks in Figures 2b and c. This trend suggests that the iodide-rich domains form by migration of iodide towards the initially formed centers, and that the bromide-rich domains form simply by depletion of iodide from the mixed phase.

Figures 2d-f show the $\Delta T/T$ during light soaking at a pressure of 0.2 GPa, other pressures are shown in Supporting Information Figures 7 to 9. We find that for an initial composition of $x = 0.25$ the perovskite is entirely stable against phase segregation at 0.2 GPa. In addition, in sharp contrast to the ambient pressure results (Figures 2b-c), the segregated iodide-rich phases of perovskites with initial composition $x = 0.5$ and $x = 0.7$ no longer occur at the same energy at 0.2 GPa. That is, $x = 0.5$ (Figure 2d) shows a clear ingrowth

of the iodide-rich phase at 1.89 eV while the $x = 0.7$ remains mostly stable, showing only a small side peak at 2.0 eV. These energies correspond to $x_s = 0.37 \pm 0.03$ (for $x = 0.5$) and $x_s = 0.51 \pm 0.04$ (for $x = 0.7$), which means that at high pressure, the segregated iodide-rich phase has a different composition depending on the initial composition (see Supporting Information Figure 10 for $\Delta T/T$ as a function of delay time). The observation that $x_s$ does vary with $x$ at high pressure (in contrast to the situation at ambient pressure, see also Supporting Information Table 1) is a first indication that $x_s$ is *not* determined by the cubic-to-tetragonal phase transition. In addition, Jaffe et al.[11] reported that both $x = 0.2$ and $x = 0.4$ are still in the cubic phase at 0.5 GPa, which means that the phase transition does not shift to higher $x$-values within this pressure range. Therefore, the cubic-to-tetragonal phase transition is not the reason for the observed shifts in the terminal $x$ at high pressure.

As an alternative explanation, we consider mechanical effects associated with the less compressible and smaller unit cell volume for samples with higher bromide content as the origin of the reduced segregation. Since $MAPbBr_3$ has a larger bulk modulus than $MAPbI_3$,[11,25] the volume change upon applying pressure to the mixed-halide $MAPb(I_{1-x}Br_x)_3$ perovskite is smaller for higher bromide contents (see Supporting Information Figure 11). Figure 3a shows $x_s$ of the iodide-rich phase (squares) and bromide-rich phase (open circles) as a function of the unit cell volume, obtained from pressure-dependent TAS measurements of segregated samples with initial stoichiometries of $x = 0.25$ (black), 0.5 (red) and 0.7 (yellow). Note that the largest volume for each $x$ corresponds to ambient pressure, and the smallest volume corresponds to 0.3 GPa. We find that the stable iodide-rich composition significantly shifts with changes in the unit cell volume $V$, increasing from the previously observed $x$~0.2 (for $V > 218$ Å$^3$) to $x$~0.6 (for $V = 209$ Å$^3$; $x = 0.7$ at 0.3 GPa).

These observations show that the unit cell volume is an important factor determining the $x$-values at which mixed-halide perovskites are stable. Indeed, we find that chemically compressing $x = 0.5$ by replacing 50% of MA with smaller Cs cations reduces the halide segregation roughly to the same extent as applying 0.2 GPa external pressure to the pure MA cation perovskite, see Figures 3b-c. Note that Cs-based perovskites have larger bandgaps than MA-based perovskites,[26] so that the absolute peak energies cannot be compared between Figure 3b and c. On fully exchanging the MA with Cs for $x = 0.5$, the bandgap is almost completely stabilized against segregation, and the energy difference between the two peaks is only ~ 100 meV, similar to applying 0.3 GPa to the MA perovskite (see Figure 3c). Whereas the pressure-dependent measurements on MAPb($I_{1-x}Br_x$)$_3$ allowed us to isolate the effect of pressure, reducing the unit cell by mixing in Cs may lead to changes in the defect density or the halide ratio. Therefore, a one-to-one relation between the physical and chemical pressure cannot be made. However, the observation that stability can be improved upon decreasing the unit cell volume, either physically or chemically, suggests that compressing the unit cell is a general route toward improved stability observed in Cs-based perovskites.[18,27]

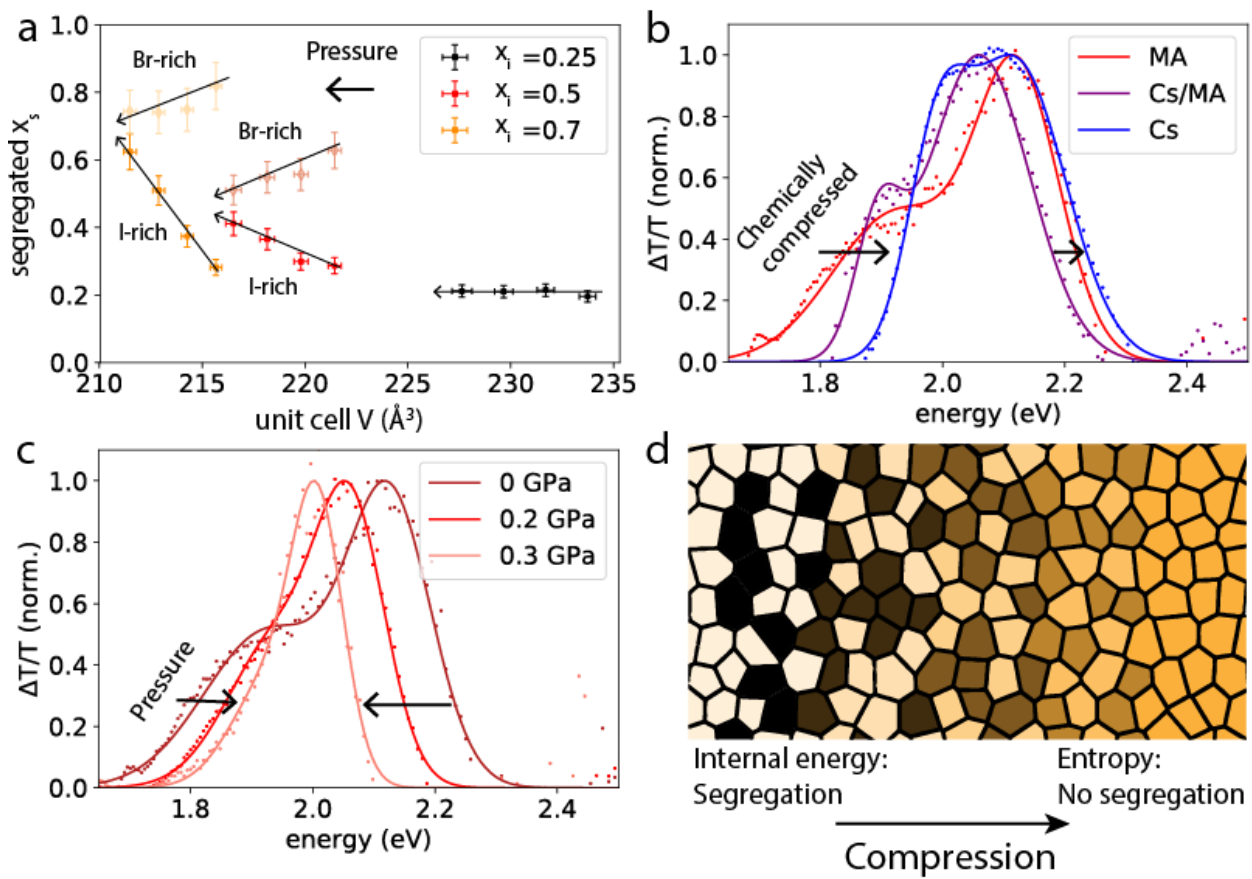

**Figure 3.** (a) Composition ($x_s$) of low-bandgap (I-rich) and high-bandgap (Br-rich) phases in MAPb($I_{1-x}Br_x$)$_3$ after segregation, plotted against initial (average) unit cell volume. (b-c) Normalized transient absorption spectra ($t_d = 15$ ps) after 20 minutes of light soaking a chemically (b) and a physically (c) compressed $x = 0.5$ mixed-halide perovskite. The samples shown in (b) correspond to MAPbI$_{1.5}$Br$_{1.5}$ (red), MA$_{0.5}$Cs$_{0.5}$PbI$_{1.5}$Br$_{1.5}$ (purple) and CsPbI$_{1.5}$Br$_{1.5}$ (blue). (d) Schematic representation of the suppression of segregation by compressing the mixed-halide perovskite. The orange color (on the right) represents the initial mixing ratio, the light- and dark-colored regions represent Br- and I-rich grains, respectively. At ambient pressure, the positive mixing enthalpy for MAPb(I$_{1-x}$Br$_x$)$_3$ drives photo-induced segregation at $x > 0.2$.[14,15] For higher pressure or Cs content, with smaller unit cell volumes, the mixing enthalpy is reduced so that the entropy dominates for a larger range of mixing ratios. As a result, the segregated mixing ratios are closer to the initial mixing ratio (in orange) in compressed perovskites.

We rationalize these observations from an alteration of $x_s$ due to the $P\Delta V$ term in the free energy of mixing (Equation 1). Whereas $\Delta S$ is unchanged under the mild pressure we apply, due to a regular perovskite structure being maintained, the previously neglected $P\Delta V$ term does change $\Delta H(x)$. As the bulk modulus of the iodide (9 GPa for MAPbI$_3$) is much smaller than the bromide (18 GPa for MAPbBr$_3$), the iodide-rich regions undergo a larger volume change and pay a larger enthalpic penalty. From continuum mechanics, the combination of $P = 0.3$ GPa and $x = 0.5$ results in $\Delta V = 5$Å$^3$, and $P\Delta V = 1$ kJ/mol. Given that the magnitude of $T\Delta S$ is limited to 2 kJ/mol, this represents a substantial contribution. In addition, the higher compressibility of iodide means that the ionic size mismatch decreases with pressure, which will lower the microscopic strain of the mixed system and allow $T\Delta S$ to dominate and stabilize the mixture for larger values of $x$ (see Figure 3d). We note that the magnitude of the terms involved here are small and are likely to result in an ensemble of ion configurations and distributions accessible around room temperature.

The observation that segregation only occurs under light requires a modification of the equilibrium thermodynamics (Equation 1)[14] in the presence of electronic excitations.[15] Local inhomogeneities in halide distribution may result in low-bandgap regions (already present in the dark),[28] which act as traps to photo-excited holes. The bandgap difference $\Delta E_g$ between the high- and low-bandgap domains then provides the driving force for the light-induced phase segregation process.[15] This electronic term in the phase diagram is expected not to change significantly with pressure, as the bandgaps of both the iodide- and bromide compounds, and their mixtures, show a similar pressure dependence (see also Supporting Information Figure 12). However, the pressure-induced shift in minimum $\Delta G$ (dark) significantly extends the miscible regime of $x$ in compressed perovskites and thus, the range of compositions that is entropically stabilized.

Previous approaches to suppress halide segregation are mainly based on the reduction of iodide vacancies, for instance by adding potassium iodide or using an excess of halides during the synthesis.[23,29] The reduction of vacancies lowers the density of mobile ions, which considerably slows ion migration and consequently the halide segregation. Slowing down the rate of halide segregation is an effective approach to *kinetically* stabilize mixed-halide perovskites.[15] However, as long as these systems are *thermodynamically* unstable, they are still prone to segregate slowly and hence, thermodynamic stabilization achieved through manipulating the unit cell is the only route to applications of mixed-halide perovskites that require long-term (*i.e.* many decades) stability.

**Conclusion**

To conclude, we have shown that compressing mixed-halide MAPb(I$_{1-x}$Br$_x$)$_3$ perovskite thin films, either *via* applying external pressure or *via* reducing the cation size, greatly improves their stability against photo-induced halide segregation. Using pressure-dependent transient absorption spectroscopy, we followed the compositional changes in both the iodide- and the bromide-rich phases associated with segregation. Whereas at ambient pressure, the segregation discontinues if the iodide-rich phase reaches $x_s \sim 0.2$, this threshold is substantially shifted with pressure, reaching $x_s \sim 0.6$ at 0.3 GPa for an initial mixing ratio of $x = 0.7$. We interpret these findings from an alteration of the Gibbs free energy *via* the $P\Delta V$ term, which was initially overlooked in the theoretical calculations. This term, which is larger than traditional inorganic semiconductors allows owing to the mechanical softness of halide perovskites, provides a lever to extend the range of thermodynamically stable mixed-halide compositions by increasing the pressure or decreasing the volume. Importantly, these results suggest that any iodide-to-bromide ratio could in principle be thermodynamically stabilized against halide segregation by tuning the crystal volume and compressibility, enabling full bandgap tunability of stable mixed-halide perovskites.

**Experimental Section**

Sample fabrication. Quartz substrates were sonicated with deionized water, acetone, and isopropanol sequentially for 15 minutes, followed by an oxygen plasma treatment for 20 minutes at 100 W. The solvents N,N-dimethylformamide (DMF, Sigma Aldrich anhydrous, ≥ 99%) and dimethylsulfoxide (DMSO, Sigma Aldrich anhydrous, ≥ 99.9%) were mixed in a 4:1 (DMF:DMSO) volume ratio. The solvent mixtures were used to prepare stock solutions of lead iodide (TCI, 99.99%, trace metals basis), $CH_3NH_3I$ (MAI, TCI, >99%), lead bromide (Sigma Aldrich, trace metals basis) and $CH_3NH_3Br$ (MABr, TCI, >98%) by dissolving these precursors at 1.1 M. $MAPbI_3$ and $MAPbBr_3$ solutions were prepared by mixing the MAI with $PbI_2$ and MABr with $PbBr_2$ stock solutions at 1:1 molar stoichiometric ratios (*i.e.* 1:1 v:v). The $MAPb(I_{1-x}Br_x)_3$ precursor solutions were prepared by mixing $x$ parts $MAPbBr_3$ stock solution with $(1 - x)$ parts $MAPbI_3$ stock solution, resulting in 1.1 M $MAPb(I_{1-x}Br_x)_3$ solutions. The films were prepared by spin-coating the precursor solutions on quartz substrates at 9,000 rpm for 30 s and anti-solvent of Chlorobenzene (Sigma Aldrich, anhydrous, ≥ 99%) was dropped 15 s after the start of spin-coating, followed by thermal annealing at 100 °C for 1 hour. Both the preparation and spin-coating of solutions were done in a nitrogen-filled glovebox.

Characterization. The XRD patterns of the perovskite films deposited on quartz were measured using an X-ray diffractometer, Bruker D2 Phaser, with Cu Kα ($\lambda$ = 1.541 Å) as X-ray source, 0.05° (2θ) as the step size, and 0.150 s as the exposure time. Elemental analysis (EDX) was performed using a FEI Verios 460 field emission scanning electron microscope (SEM) operated at 7 kV. Pressure-dependent transmission spectra of the $MAPb(I_{1-x}Br_x)_3$ films were measured from 850 nm to 350 nm using a pressure cell (ISS Inc.) and a LAMBDA 750 UV/Vis/NIR Spectrophotometer (Perkin Elmer). The absorption at ambient pressure was determined by measuring the transmission and reflection inside the integrating sphere.

Transient absorption spectroscopy under hydrostatic pressure. During the measurement the sample was kept inside a high-pressure cell (ISS Inc.) filled with the inert pressurizing liquid, Fluorinert FC-72 (3M). Prior to use, the liquid was degassed in a Schlenk line by bubbling nitrogen to remove oxygen and water. The hydrostatic pressure was generated through increasing the amount of the pressurizing liquid in the cell using a manual pump. The pressure was increased from ambient to 0.3 GPa in steps of 0.1 GPa, with an error of 0.02 GPa estimated from reading the pressure on the gauge (unless at ambient pressure). At every pressure, we waited 7 minutes for equilibration of the material under pressure. Transient absorption (TA) spectra were collected using a home-built set-up operating in a non-degenerate pump-probe configuration. The laser source for the TA is a regenerative Ti:sapphire amplifier (Coherent) producing a fundamental beam characterized by 800 nm pulses at a 1 kHz repetition rate with a pulse duration of 35 fs and a pulse energy of 6.5 mJ. The fundamental beam is split in two beams by a beam splitter. After chopping the beam in the pump path (using 500 Hz as frequency), 400 nm pulse pump was generated by doubling 800 nm pulse with a beta barium borate (BBO) crystal. A short-pass filter was placed after the BBO crystal in the pump path, to remove 800 nm residue from the fundamental beam. The white-light continuum probe pulses were produced by focusing the 800 nm fs pulses through a 2 mm-sapphire plate. The probe spot size was chosen to be smaller than the pump spot size to obtain homogenous excitation over the probed area. The two beams were then spatially overlapped inside the pressure cell. To follow the evolution in time of the system, the pump-probe delay time was changed from 0 to 1000 ps using a mechanical delay stage. Unless states otherwise, the pump excitation density was ~$10^{18}$ cm$^{-3}$ (see Supporting Information Figure 13 for different excitation densities). The samples were light-soaked using a 405 nm continuous wave (cw) single-mode fiber-coupled laser source (Thorlabs). The cw laser was focussed on the same spot as the pump and probe and its spot size (diameter of 243 micrometer) was large enough to fully cover the probed area (< 50 micrometer). The intensity of the light-soaking source was $2.37 \times 10^3$ mW/cm$^2$ (~24 sun), and no segregation was observed in absence of the cw source (see Supporting Information Figure 14).

## Author Information

The authors declare no competing interests.

## Author contributions

E.M.H. conceived the idea and performed the pressure-dependent TA experiments and data analysis together with L.A.M., under the supervision of B.E. F.W. and L. McG. assisted in the sample preparation and J.V. assisted in the TA experiments under the supervision of H.J.B. Thermodynamic calculations were performed by A.W, Y. W. and Y. J. The manuscript was written by E.M.H. and A.W. with input from all other authors.

**Acknowledgments**

The work of E.M.H., L.A.M., F.W., J.V., L. McG., H.B. and B.E. is part of the Dutch Research Council (NWO) and was performed at the research institute AMOLF. This research was also supported by the Creative Materials Discovery Program through the National Research Foundation of Korea (NRF) funded by Ministry of Science and ICT (2018M3D1A1058536). We are grateful to the UK Materials and Molecular Modelling Hub for computational resources, which is partially funded by EPSRC (EP/P020194/1). The authors thank Henk-Jan Boluijt for the design of Figure 1a.